\documentclass[]{emulateapj}
\newcommand{\ha}{H$\alpha$}
\newcommand{\hb}{H$\beta$}
\newcommand{\hg}{H$\gamma$}

\received{2007 February 9}
\begin{document} 

\title{New Close Binary Systems from the SDSS--I (Data Release Five)
and the Search for Magnetic White Dwarfs in Cataclysmic Variable
Progenitor Systems}
 
\shorttitle{Magnetic WDs in CV Progenitors} 
\shortauthors{Silvestri et~al.}

\author{Nicole M. Silvestri\altaffilmark{1}, Mara
  P. Lemagie\altaffilmark{1}, Suzanne L. Hawley\altaffilmark{1},
  Andrew A. West\altaffilmark{2}, Gary D. Schmidt\altaffilmark{3}, James
  Liebert\altaffilmark{3}, Paula Szkody\altaffilmark{1}, Lee
  Mannikko\altaffilmark{1}, Michael A. Wolfe\altaffilmark{1},
  J. C. Barentine\altaffilmark{4}, Howard
  J. Brewington\altaffilmark{4}, Michael Harvanek\altaffilmark{4},
  Jurik Krzesinski\altaffilmark{5}, Dan Long\altaffilmark{4}, Donald
  P. Schneider\altaffilmark{6}, and Stephanie
  A. Snedden\altaffilmark{4}}

\altaffiltext{1}{Department of Astronomy, University of Washington,
  Box 351580, Seattle, WA 98195, USA; nms@astro.washington.edu,
  mlemagie@u.washington.edu, slh@astro.washington.edu,
  szkody@astro.washington.edu, leeman@u.washington.edu,
  maw2323@u.washington.edu.}

\altaffiltext{2}{Astronomy Department, 601 Campbell Hall, University
  of California, Berkeley, CA 94720, USA; awest@astro.berkeley.edu}

\altaffiltext{3}{Department of Astronomy and Steward Observatory,
  University of Arizona, Tucson, AZ 85721, USA;
  schmidt@as.arizona.edu, jliebert@as.arizona.edu.}

\altaffiltext{4}{Apache Point Observatory, P.O. Box 59, Sunspot, NM
  88349, USA; jcb@apo.nmsu.edu, hbrewington@apo.nmsu.edu,
  harvanek@apo.nmsu.edu, long@apo.nmsu.edu, snedden@apo.nmsu.edu.}

\altaffiltext{5}{Mt. Suhora Observatory, Cracow Pedagogical University,
  ul. Podchorazych 2, 30-084 Cracow, Poland; jurek@apo.nmsu.edu.}

\altaffiltext{6}{Department of Astronomy, Penn State University, PA
  16802 USA; dps@astro.psu.edu}

\begin{abstract}

We present the latest catalog of more than 1200
spectroscopically--selected close binary systems observed with the
Sloan Digital Sky Survey through Data Release Five.  We
use the catalog to search for magnetic white dwarfs in cataclysmic
variable progenitor systems.  Given that approximately 25\% of
cataclysmic variables contain a magnetic white dwarf, and that our
large sample of close binary systems should contain many progenitors
of cataclysmic variables, it is quite surprising that we find only two
potential magnetic white dwarfs in this sample.  The candidate
magnetic white dwarfs, if confirmed, would possess relatively low
magnetic field strengths ($B_{\rm WD} < 10$ MG) that are similar to
those of intermediate--Polars but are much less than the average field
strength of the current Polar population.  Additional observations of
these systems are required to definitively cast the white dwarfs as
magnetic.  Even if these two systems prove to be the first evidence of
detached magnetic white dwarf + M dwarf binaries, there is still a
large disparity between the properties of the presently known
cataclysmic variable population and the presumed close binary
progenitors.

\end{abstract}



\section{Introduction \label{sec-intro}}

The evolution of stars in close binary systems leads to interesting
stellar end-products such as cataclysmic variables (CVs), Type 1a
supernovae, and helium--core white dwarfs (WDs). The period in which
an evolved star ascends the asymptotic giant branch and engulfs a
close companion in its evolving atmosphere, referred to as the common
envelope phase, probably plays a dominant role in the evolution of
these systems and as yet is poorly understood.  The angular momentum
of the system is believed to aid in the eventual ejection of the
common envelope to reveal the remnant WD and close companion.  After
the common envelope has been ejected, gravitational and magnetic
braking work to decrease the orbital separation of the detached system
\citep{deKool93}.  This orbital evolution continues through to the
CV phase. The effect of the common envelope on the secondary star in
these systems is another aspect of close binary evolution which is not
well characterized. Plausible scenarios for the secondary companion
range from accreting as much as 90\% of its mass during this phase to
escaping relatively unscathed from the common envelope, emerging in
the same state as it entered \citep[see][and references
therein]{Livio96}.

Recently, studies of close binary systems with WD companions
\citep[see for example][]{Farihi05a,Farihi05,Pourbaix05,S06} have
revealed yet another puzzling property of these systems.  None of the
WDs in close binary systems with low--mass, main sequence companions
appear to be magnetic \citep{Liebert05}.  Close, non--interacting
binary systems with WD primaries are quite common and are believed to
be the direct progenitors to CVs \citep[][and references
therein]{Langer00}.  Magnetic WDs, stellar remnants with magnetic
fields in excess of $\sim 1$ MG, comprise only a small percentage of
the isolated WD population \citep[$\sim 2$\%,][]{Liebert05}.  Note
that the 2\% magnetic WD fraction applies to magnitude--limited
samples like the Palomar--Green \citep{Liebert80}.  However, the same
paper notes that magnetic WDs may generally have smaller radii than
non--magnetic white dwarfs, due to higher mass.  In a given volume,
the density of magnetic WDs may be $\sim 10$\% of all WDs
\citep{Liebert03}. The SDSS is also a magnitude limited sample so we
assume a similar expected value for the close binaries. Our sample (as
discussed in detail in \S2) contains 1253 potential close binary
systems.  Therefore we assume approximately 24 of these binaries to
harbor a magnetic WD.  Possible implications of the small radii for
magnetic WD + main sequence pairs will be discussed in
\S5.  However, more than 25\% of the WDs in the
currently identified CV population are classified as magnetic, and
many have magnetic fields in excess of 10 MG \citep[see][]{WF00}.

\citet{Holberg02} have compiled a list of 109 known WDs within 20pc
(and complete to within 13pc) that have nearly complete information
about the presence of a companion.  Of the 109 WDs in their sample,
$19 \pm 4$ have nondegenerate companions.  Table~7 in \citet{Kawka06}
lists all known magnetic WDs as of June 2006.  Of the magnetic WDs
listed in their table, 149 have field strengths identifiable in
SDSS-resolution spectra ($B_{WD} \geq 3$ MG).  If the magnetic WDs in
the \citet{Kawka06} sample are assumed to be drawn from a similar
sample then $28 \pm 5.3$ would be expected to have nondegenerate
companions, and yet none have been detected in the \citet{Kawka06}
sample.  This is nearly a $5\sigma$ deficit in magnetic WDs with
nondegenerate companions.

\citet{HM05} looked at the 2MASS $JHK_{s}$ photometry of 347 WDs in
the Palomar--Green sample.  Of the 347 WDs, 254 had reliable
infrared measurements of at least $J$ magnitude.  Of these, 59 had
excesses indicative of a nondegenerate companion and another 15 showed
``probable'' excesses \citep{Liebert05}.  This gives a WD+dM fraction
of 23\% (definite excess) and 29\% (including all probable
excesses). If the \citet{Kawka06} sample had the same frequency of of
nondegenerate companions as the Palomar--Green sample, they should
have 34 and 43, respectively.  This is nearly as $6\sigma$ deficit!

This apparent lack of magnetic WDs with main sequence companions is
not restricted to studies of close binaries.  Low resolution
spectroscopic surveys of more than 500 common proper motion binary
systems discovered by \citet{Luyten64,Luyten68,Luyten72} and
\citet{GBT71,GBT78} revealed no magnetic WDs paired with main sequence
companions in these wide pairs \citep{Smith97,S05}. In addition,
\citet{Schmidt03} and \citet{Vanlandingham05} have identified over 100
magnetic WDs in the Sloan Digital Sky Survey
\citep[SDSS,][]{Gunn98,York00,Stoughton02,Pier03,Gunn06}.  As
discussed by \citet{Liebert05}, this implies essentially no overlap
between the close binary and magnetic WD samples.

A new class of short--period, low accretion--rate polars (LARPS)
identified by \citet{Schmidt05b} may explain, in part, these
``missing'' magnetic WD systems.  In these systems, the donor star has
not filled its Roche Lobe.  The WD accretes material by capturing the
stellar wind of the secondary.  These CVs have accretion rates that
are less than 1\% of accretion rates normally associated with CVs.
The discovery of these systems sheds some light on the whereabouts of
magnetic WD binaries, though as \citet{Schmidt05b} point out, this
still does not explain the apparent lack of long--period, detached
magnetic WD systems.  Thought to be the first detached binary with a
magnetic WD, SDSS J121209.31+013627.7, a magnetic WD with a probable
brown dwarf (L dwarf) companion \citep{Schmidt05a} has been shown to
be one of these LARP systems \citep{Debes06,KM06,Burleigh06}.  To
date, magnetic WDs have only been found as isolated objects, in
binaries with another degenerate object (WD or neutron star
companion), or in CVs; none have a clearly main sequence
companion.

In this study, we investigate a new large sample of close binary
systems in an effort to uncover these ``missing'' magnetic WD binary
systems.  The sample comprises more than 1200 close binary systems
containing a WD and main sequence star drawn from the SDSS, many of
which were originally presented in \citet[][hereafter, S06]{S06}. We
find that \emph{only two} of the WDs in these pairs appear to be
magnetic. Even if confirmed, neither of these WDs has magnetic field
strength comparable to those observed in the majority of magnetic
(Polar) CV systems.  We confirm that the current CV and close binary
populations are indeed disparate and show that more work is necessary
to unravel this mystery.

In \S2 we introduce the catalog of close binary systems
through the public SDSS Data Release Five \citep[DR5;][]{AMDR5}.  We
discuss our analysis techniques in \S3 and we present
our results in \S2.  Our discussion and concluding
remarks are given in \S5 and \S6,
respectively.

\section{The SDSS Close Binary Catalog through DR5 \label{sec-cat}}

The combined properties of the majority of close binaries in this
paper are discussed in detail in \citet{Raymond03} and S06.  The S06
catalog was based on a preliminary list of spectroscopic plates
released internally to the collaboration and as such does not include
objects from $\sim 200$ plates released with the final public Data
Release Four \citep[DR4;][]{AMDR4}.  The additional systems from both
DR4 and DR5 do not change the overall results from analysis performed
in S06, hence no new analysis is presented here. We include this list
in its entirety to complete the DR4 catalog introduced by S06 and add
over 300 new systems from the now public DR5 \citep{AMDR5}.  This
completes the catalog of close binary systems with a WD identified
through SDSS--I.  More close binaries are being targeted in the
SDSS--II (SEGUE) survey which will continue to increase the sample
through 2008.

The list of 1253 potential close binary systems given in
Table~\ref{wdmcat} includes objects from all plates released with the
public DR5, thereby superseding the S06 DR4 catalog. The technique
used to search for these objects is the same as described in S06.  As
with that study, we do not include systems with low signal-to-noise
ratios (S/N $< 5$) and do not search for systems with non--DA WDs.  We
emphasize that our sample is not complete (or bias free) due to the
selection effects imposed by our detection methods and due to the
sporadic targeting of these objects in the SDSS spectroscopic survey
as discussed in S06.  Thus, our sample represents primarily bright,
DA~WD~+~M dwarf binary systems.  As evidenced by \citet{Smolcic04},
there are potentially thousands more WD~+~M dwarf binaries observed
photometrically in the SDSS but not targeted for spectroscopy.  Our
catalog represents an interesting and statistically significant
sampling of these systems, the properties of which can be used to test
models of close binary evolution \citep[see][for example]{PW06}.

The list of plate numbers from which this sample has been drawn can be
found at http://das.sdss.org/DR5/data/spectro/1d\_23/.  This plate
list includes both ``extra'' and ``special'' plates. The extra plates
are repeat observations of survey plates taken during normal
operation.  The special plates are observations for special programs
(e.g. SEGUE, F~stars, main sequence turnoff stars, quasar selection
efficiency, etc.) that are not part of the original SDSS--I survey.

The first four columns of Table~\ref{wdmcat} list the SDSS identifier,
the plate number, fiber identification, and modified Julian date (MJD)
of the observation, followed by the spectral type of the components
(determined visually) where Sp1 represents the blue object and Sp2 is
the red object.  Columns~6 and 7 give the J2000 coordinates (in
decimal degrees) for the object.  The next 15 columns give the $ugriz$
PSF photometry \citep{Fukugita96,Hogg01,Ivezic04,Smith02,Tucker06},
photometric uncertainties ($\sigma_{ugriz}$), and reddening
($A_{ugriz}$).  The magnitudes are not corrected for Galactic
extinction. Column~23 lists the SDSS data release in which the object
was discovered as well as additional references in the literature.
Additional notes for the objects are listed in column~24.

\begin{figure}
\epsscale{0.50}
\plotone{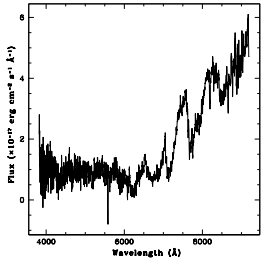}
\caption{Example of an M dwarf with excess blue flux (:+dM) from 
  Table~\ref{wdmcat}.  The companion is seen as little more than excess 
  blue flux in the M dwarf spectrum.  Follow-up spectroscopy to resolve 
  the companion is necessary to rule out the presence of a magnetic WD.  
  Note: spectrum has been boxcar smoothed with a filter size of
  seven. \label{bluedM}} 
\end{figure}

The objects identified in Table~\ref{wdmcat} as :+dM are likely M
dwarfs with faint, cool WD companions.  The discovery spectra for
these objects reveal little more than excess blue flux at wavelengths
shorter than 5000 \AA, as shown in Figure~\ref{bluedM}.  It is
possible that some of these pairs may contain a magnetic WD; however,
much higher S/N spectra are required to adequately characterize the blue
component of these systems.

Similarly, the thirty nine objects identified as WD+: or WD+:e (see
Figure~8 of S06) have either some excess flux in the red or have
emission at Balmer wavelengths indicative of a faint, active,
low--mass or sub--stellar companion. The companion to the magnetic WD
in \citet{Schmidt05a} was first identified by emission at \ha\ in the
SDSS discovery spectrum.  Other than the emission at \ha\ this object
had no other optical signature of a companion.  We are performing
followup observations using the ARC 3.5--m telescope at Apache Point
Observatory to obtain radial velocities and near--infrared imaging of
these objects to measure the orbital periods and categorize the
probable low--mass companion's spectral type. We have already
confirmed that none of these systems contain a magnetic WD.


\section{The Search for Magnetic WDs \label{sec-search}}

\begin{figure}
\epsscale{0.50}
\plotone{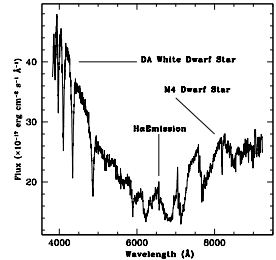}
\caption{A Typical WD+dM System: SDSS J140723.03+003841.7, the
superposition of a DA (hydrogen atmosphere) WD and a M4 red dwarf
star. \ha\ emission is visible in many of these systems and is a
result of chromspheric activity on the surface of the M star, perhaps
enhanced due to the influence of the WD. The lack of broad Zeeman
absorption features in the hydrogen lines indicates that the magnetic
field strength of the WD is very low (compare with
Figure~\ref{fake}). \label{example}}
\end{figure}

\citet{Schmidt03} and \citet{Vanlandingham05} demonstrated that
magnetic WDs with field strengths as low as $\sim3$ MG can be
effectively measured using SDSS spectra.  Visual inspection of the
systems in our sample reveals no obvious magnetic WDs in spectra with
good S/N ($> 10$) \citep{Lemagie04}.  Most are classical WD~+~M dwarf close
binaries as shown in Figure~\ref{example}.  Of interest are the lower
quality spectra, where the features of the WD are less obvious because of
low S/N and/or contamination by the spectral features of the close M dwarf
companion.  These effects make it difficult to identify small
magnetic field effects on the WD absorption features.  Thus,
relatively low magnetic fields ($B_{\rm WD}< 10$ MG) are not easily
recognized in the combined spectrum.

\subsection{The Simulated Magnetic Binary Systems}

Given the difficulties associated with visually identifying features
in these systems, we developed a method to search for the
characteristic Zeeman splitting of the DA WD absorption features that
is also sensitive to low magnetic field WDs.  We use a program that
attempts to match absorption features in magnetic DA WD models
\citep[see][and references therein for details on the
models]{Kemic74a,Kemic74b,Schmidt03} through an iterative method of
smoothing and searching the stellar spectrum. To develop a robust
program to search for magnetic WDs in close binaries we first tested
our program on WDs of known magnetic field strength.  We used the
magnetic DA WDs with field strengths between 1.5 MG $\leq B_{\rm WD}
\leq 30$ MG from \citet{Schmidt03} and \citet{Vanlandingham05} as our
test sample.  We then constructed model spectra at every half--MG
between 1.5 MG $\leq B_{\rm WD} \leq 30$ MG, each with magnetic field
inclinations of 30\degr, 60\degr, and 90\degr.  The program was able
to match (using a $\chi^{2}$ minimization) the magnetic field strength
of each of the magnetic WDs to within $\pm 5$ MG of the value quoted
in \citet{Schmidt03}.

\begin{figure}
\epsscale{0.50}
\plotone{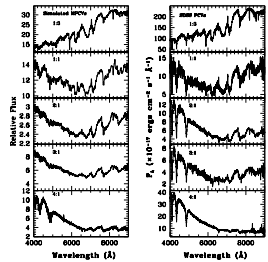}
\caption{Comparison of simulated and observed pre--cataclysmic
  variable (PCV) systems.  Left Hand Column: Simulated magnetic PCVs
  produced by adding WD spectra from \citet{Schmidt03} to M dwarf
  spectra from \citet{Hawley02} with brightness ratios as specified at
  6500 \AA. Right Hand Column: Observed PCVs from
  \citet{S06}. \label{comp}}
\end{figure}

We then constructed a sample of simulated SDSS spectra of magnetic
binary systems. The simulated binaries were created by adding the
spectra of magnetic WDs used in our initial test from
\citet{Schmidt03} and \citet{Vanlandingham05} to the M star templates
of \citet{Hawley02}.  We first normalized all spectra at a wavelength
of 6500 \AA, and then combined them with flux ratios of 1:4 (WD:M
dwarf) to 4:1 to replicate the range of flux ratios observed in the
close binary sample (see Figure~\ref{comp})\footnote{Note that 6500
\AA\ is the midpoint of the SDSS combined blue and red spectra, as
plotted in Figure~\ref{comp}.  In reality the SDSS spectra extend to
below 3900 \AA\ and to nearly 10000 \AA.}. This created a sample of
binaries which represent the average brightness and spectral type
distribution of the majority of the systems in Table~\ref{wdmcat}
(i.e. DA WDs and M0--M5 dwarfs).

\begin{figure}
\epsscale{0.50}
\plotone{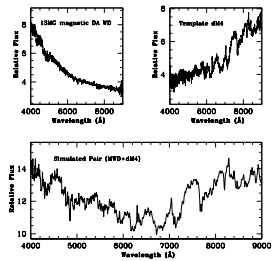}
\caption{A Simulated System. Top Left Panel: A 13 MG magnetic WD from
  \citet{Schmidt03}. Top Right Panel: Template M4 dwarf star from
  \citet{Hawley02}. Bottom Panel: addition of the magnetic WD and template
  M dwarf, assuming equal flux density at 6500 \AA.  \label{fake}}
\end{figure}

Figure~\ref{fake} is an example of one of the simulated magnetic
binary systems.  The upper left hand panel is the SDSS spectrum of a
13 MG magnetic WD, the upper right hand panel is the spectrum of a
template M4 dwarf star.  The bottom panel is the addition
(superposition) of the two spectra with a flux ratio of 1:1 at 6500
\AA.  As shown, this WD with a relatively moderate magnetic field,
when combined with the spectrum of an average M dwarf, is clearly
detected at the resolution of the SDSS spectra (R $\sim 1800$).  

\subsection{Results from the Simulated Systems \label{subsec-results}}

We found that detecting the presence of a WD magnetic field depends
most strongly on the spectral type and relative flux of the M dwarf
companion.  Due to the selection effects of the close binary sample
(see S06 for details), the majority of the M dwarfs in these binaries
have spectral sub--types between M0--M4.  In SDSS spectra, early M
dwarf spectral types contribute nearly as much flux in the blue
portion of the spectrum (4000--7000 \AA) as they do in the red
(7000--10000 \AA).  The spectrum of the blue magnetic WD is then
superimposed onto the numerous blue molecular features of the M dwarf.
This makes the small absorption features stemming from the subtle
influence of a weak magnetic field difficult to detect.

\begin{figure}
\epsscale{0.50}
\plotone{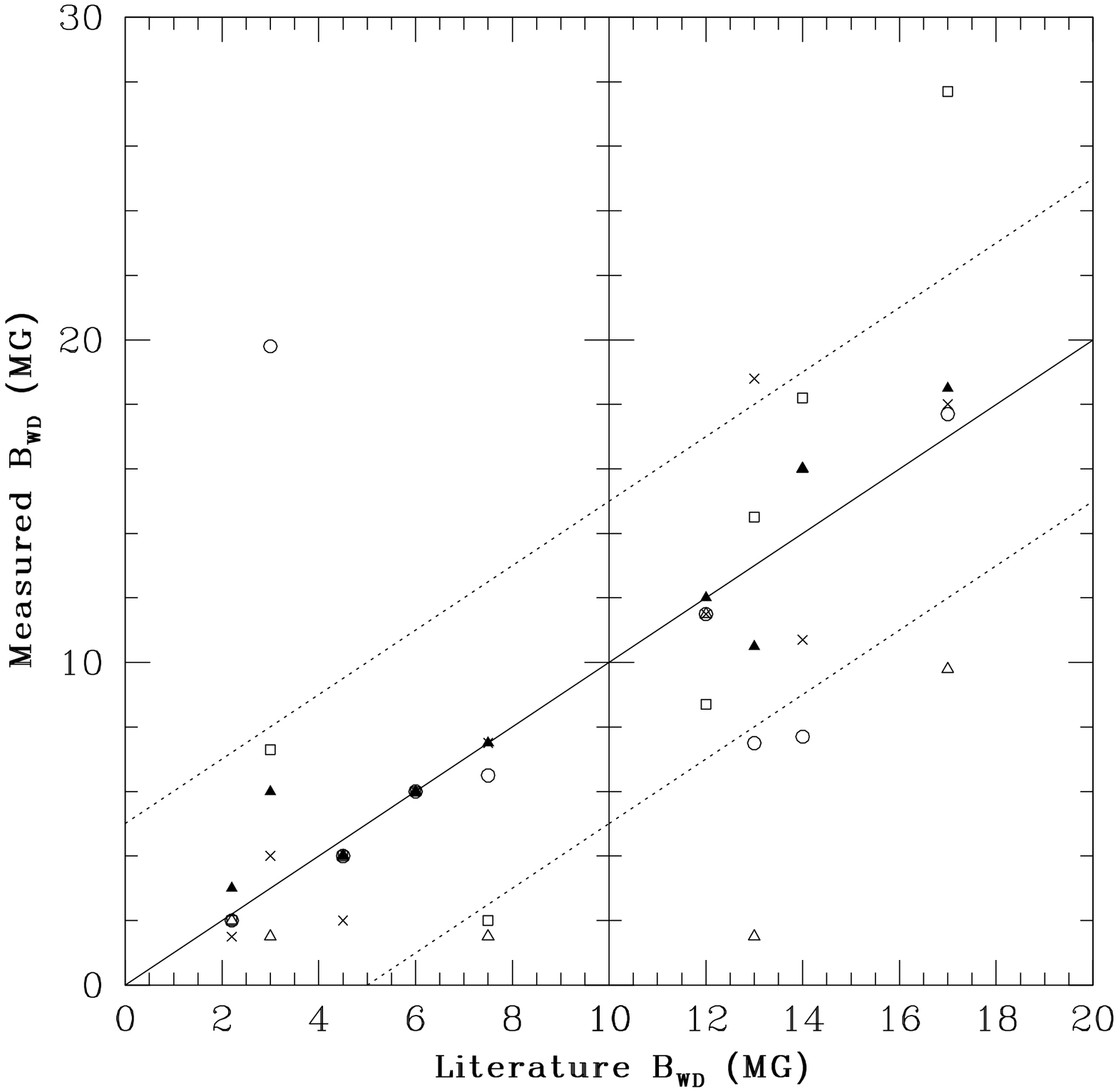}
\caption{Left Hand Panel: Subset of the simulated binary systems
  comprised of early--M dwarfs from \citet{Hawley02} paired with magnetic
  WDs and literature values from \citet{Schmidt03} and
  \citet{Vanlandingham05} with $B_{\rm WD} \leq 10$ MG.  Right Hand
  Panel: Same M dwarfs from Left Panel paired with magnetic WDs with
  $B_{\rm WD} \geq 10$ MG. The measured values are from our program.
  In both panels, the filled triangles represent single WDs, open
  squares are WD+M0, open circles are WD+M1, open triangles are WD+M2,
  and crosses are WD+M3.  The solid line has a slope of one and the
  dashed lines are $\pm 5$ MG.  Refer to \S~\ref{subsec-results} of
  the text for details. \label{earlymwd}}
\end{figure}

\begin{figure}
\epsscale{0.50}
\plotone{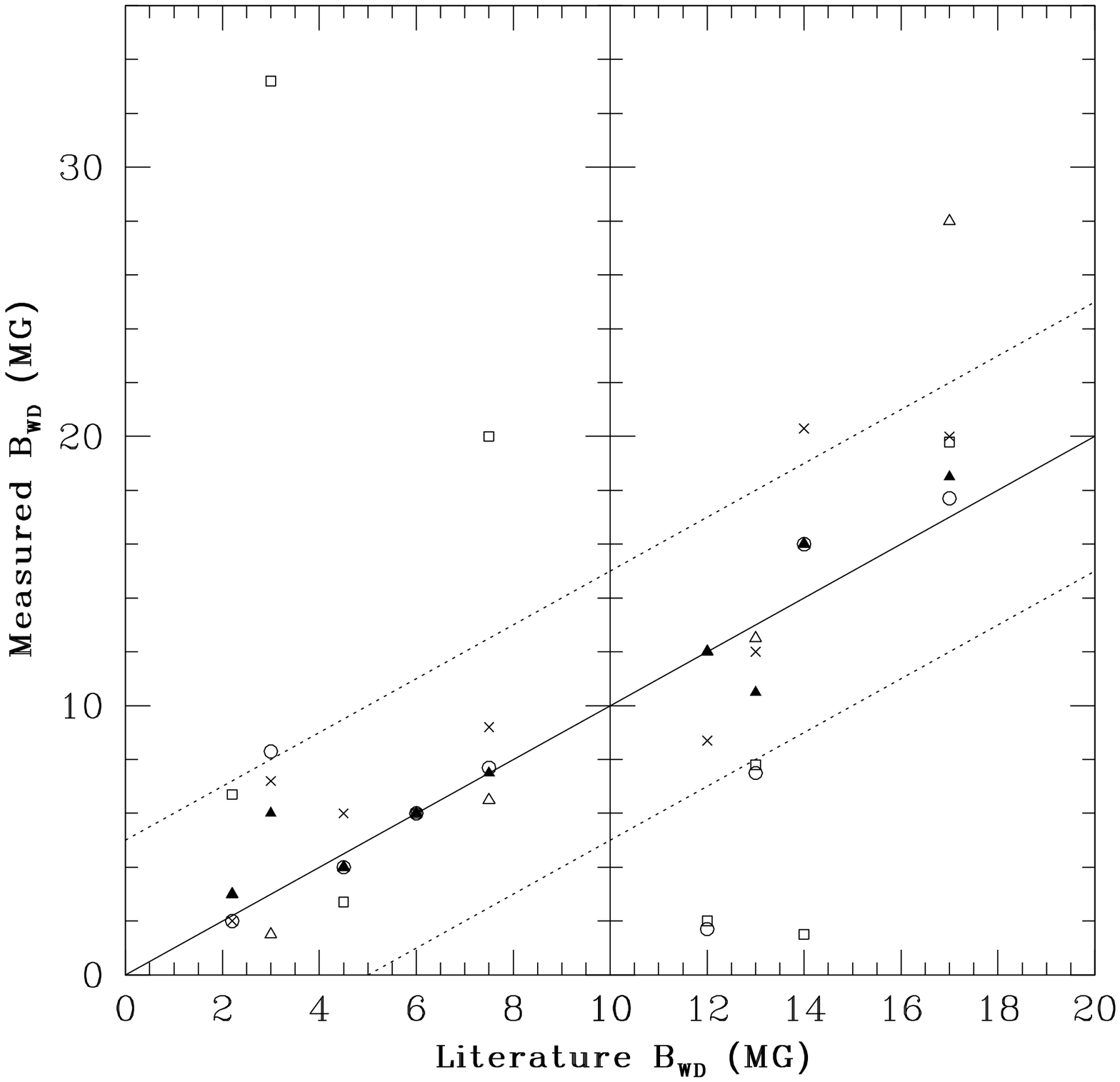}
\caption{Left Hand Panel: Subset of the simulated binary systems
  comprised of late-M dwarfs from \citet{Hawley02} paired with magnetic WDs
  and literature values from \citet{Schmidt03} and
  \citet{Vanlandingham05} with $B_{\rm WD} \leq 10$ MG.  Right Hand
  Panel: Same M dwarfs from Left Panel paired with magnetic WDs with
  $B_{\rm WD} \geq 10$ MG. The measured values are from our program.
  In both panels, the filled triangles represent single WDs, open
  squares are WD+M4, open circles are WD+M5, open triangles are WD+M6,
  and crosses are WD+M7.  The solid line has a slope of one and the
  dashed lines are $\pm 5$ MG.  Refer to \S~\ref{subsec-results} of
  the text for details. \label{latemwd}}
\end{figure}

We plot a subset of our simulated pairs to demonstrate some of these
issues in Figure~\ref{earlymwd} and Figure~\ref{latemwd}. In
Figure~\ref{earlymwd} we selected four early--type template M dwarfs
(WD+M0 = open squares, WD+M1 = open circles, WD+M2 = open triangles,
and WD+M3 = crosses) from \citet{Hawley02} and added them to a range of
magnetic WDs from \citet{Schmidt03} and \citet{Vanlandingham05}.  The
quoted value from \citet{Schmidt03} for the magnetic field strength of
each of these WDs represents the ``Literature $B_{\rm WD}$'' value on
the x--axis.  The ``Measured $B_{\rm WD}$'' is the value returned by
the program. Values returned by the program that matched the
literature values fall along the solid line.  The dashed lines
represent $\pm 5$ MG of the literature value. Figure~\ref{latemwd} is
the same except we add the same magnetic WDs to later--type M dwarf
templates (WD+M4 = open squares, WD+M5 = open circles, WD+M6 = open
triangles, and WD+M7 = crosses).  The solid triangles represent the
tests using the isolated WD spectra.

In both Figures~\ref{earlymwd} and \ref{latemwd} the program returns
the value of the single WD to within $\sim \pm 2$ MG for the large
majority of the systems.  The uncertainty of the fitted value and the
spread in values increases for magnetic fields of 3 MG or less when
the magnetic WD is paired with an M dwarf of comparable brightness.
The flux minima associated with the Zeeman features for such low
field strengths are just barely resolvable in high S/N spectra of
isolated SDSS WDs \citep[see][]{Schmidt03}.  The added complexity of
the M dwarf molecular features and the generally lower S/N spectra make it
difficult to measure the magnetic features for low magnetic field
strengths.  However, WDs with magnetic fields $\geq 4$ MG were easily
measured at all M dwarf spectral types.

In both Figures, the largest discrepancies between the literature and
measured values occur when the WD's magnetic field is between 12 MG
$\leq B_{\rm WD} \leq 18$ MG; this is true when the WD is paired with
both early-- and late--type M dwarfs.  Inspection of the model results
indicates that at these field strengths, the Zeeman features overlap
on wavelengths with strong M dwarf molecular features, causing
confusion in the identification of the feature.  However, WD spectra
with these and larger field strengths are quite easily recognized
visually so we are confident that no systems with $\geq 10$ MG have
escaped notice, though the exact value of the field strength
would be more uncertain.

\begin{figure}
\epsscale{0.50}
\plotone{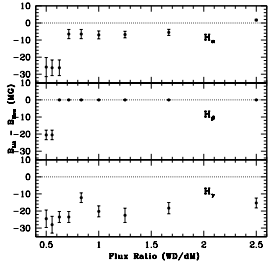}
\caption{ Here, we plot the flux ratio (WD flux/ M dwarf [dM]) versus
  the difference between the literature value
  \citep[from][]{Schmidt03,Vanlandingham05} of the magnetic field
  strength ($B_{\rm Lit}$) and the measured magnetic field strength
  ($B_{\rm Mea}$) as determined from the WD \ha\ (top panel), \hb\
  (center panel), and \hg\ (bottom panel) absorption features.  Error
  bars are from the $\chi^{2}$ fit. Refer to \S~\ref{subsec-results}
  of the text for details.
  \label{abg}}
\end{figure}

In Figure~\ref{abg}, we demonstrate the effect of the relative flux
ratio (WD: M dwarf [dM]) on the identification of the magnetic field
strength of WDs in the simulated binary sample.  The Figure gives the
relative flux ratio versus the difference between the magnetic fields
quoted in the literature and those returned by the program. We use the
same $B_{\rm WD}$ distribution in Figure~\ref{abg} as used in
Figure~\ref{earlymwd} and Figure~\ref{latemwd}.  The literature values
($B_{\rm Lit}$) are from \citet{Schmidt03} and \citet{Vanlandingham05}.
The three panels show ratios determined using \ha\ (top), \hb\
(center), and \hg\ (bottom). The program consistently returns the
quoted $B_{\rm WD}$ as determined from \hb\ until the flux
contribution from the M dwarf is nearly double the flux contribution
from the WD.  The program returns the magnetic field from the \ha\
feature to within $\pm 5$ MG until the flux contribution from the M
dwarf is nearly 1.5$\times$ the flux from the WD.  The $B_{\rm WD}$ as
measured by \hg\ is consistently 15--25 MG larger than the $B_{\rm WD}$
value in the literature at any flux ratio. The contribution of a
relatively clean spectral region near \hb, together with the fairly
strong Zeeman signal at this wavelength makes \hb\ a reliable
indicator of WD magnetic field strength for binaries with flux ratios
up to 1:2.

\section{Two Possible Magnetic WDs in the DR5 Close Binary Sample 
\label{sec-two}}

The method employed by S06 to split the binary system into its two
component spectra through an iterative method of fitting and
subtracting WD model atmospheres and template M dwarf spectra was not
used on these objects.  There are no obviously strong magnetic WDs in
the sample, suggesting that any possibly magnetic WDs must possess
relatively weak fields.  The subsequent fitting and
subtraction of model WDs and template M dwarfs adds noise to the
spectrum which would make detection of an already weak magnetic field
even more difficult.  Also, we would be subtracting a non--magnetic WD
model from the spectrum of a potentially magnetic WD in our attempt to
improve the M dwarf template fit. This adds absorption features where
none actually exist, further corrupting the WD spectrum.  Given these
complications, we chose to work with the original composite SDSS
discovery spectra.

Table~\ref{maybemags} lists the properties of the only two close
binary systems flagged by our program as containing potential magnetic
WDs: SDSS J082828.18+471737.9 and SDSS J125250.03$-$020608.1. The
first four columns are the same as for Table~\ref{wdmcat}, followed by
the R.A. and Decl. (J2000 coordinates). The tentative magnetic field
strengths (in MG), inclination of the WD magnetic field to the line of
sight (in degrees) and the spectral types of the components are listed
in Columns~7--9. For each of these systems the magnetic field strength
estimate is based upon a match to at least two of the three Balmer
features (\ha, \hb, and/or \hg) to within $\pm 5$ MG of the model
minima. The last six columns give the $ugriz$ photometry and the SDSS
data release for the objects.  Refer to Table~\ref{wdmcat} for a full
listing of photometric errors, reddening and alternate literature
sources.

\begin{figure}
\epsscale{0.50}
\plotone{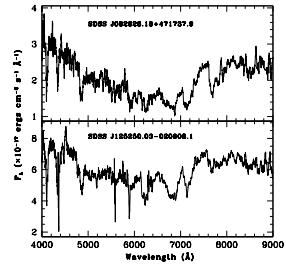}
\caption{Two potential magnetic DA WD + M dwarf pairs as identified by
our program. The tentative magnetic field strengths are 8 MG $\pm 5$
MG (top) and 3 MG +5/$-$3 MG (bottom) as determined from the \ha\ and
\hb\ WD absorption features. \label{candidates}}
\end{figure}

Figure~\ref{candidates} displays the spectra of these two objects,
which have relatively low S/N ($\sim 5$ at \ha) The identification of
the magnetic field strength was determined from the \ha\ and \hb\
features in each spectrum, which upon closer inspection may show some
Zeeman splitting. The best fit model for SDSS J082828.18+471737.9 has
a magnetic field strength of 8 MG and an inclination of 90\degr, while
the best fit model for SDSS J125250.03$-$020608.1 has a magnetic field
strength of 3 MG and an inclination of 90\degr. \hb\ appears to be
distorted in both systems, indicating a potential broadening of a few
MG field, however \hg\ and H$_{\delta}$ would show more splitting
than \hb\, but both appear to be relatively sharp in comparison.  \hb\
may be affected by TiO features from the M dwarf and there does appear
to be a minor glitch in the blue portion of the spectrum, indication
difficulty with SDSS flux calibration.  

\section{Discussion \label{sec-dis}}

Of the 1253 potential close binary systems in the DR5 catalog, there
were 168 systems that we could not measure with our program.  These
include the :+dM systems and binaries with non-DA WDs. We were not
able to unambiguously determine if the :+dM systems have a magnetic or
a non--magnetic WD as the blue component is barely visible in most of
the :+dM SDSS spectra.  Until we can identify the companion, we can
not make any statement about magnetism in these objects. The :+dM
cases where a blue component is seen in the spectrum which must be a
WD, but too faint even to classify the type may include (a) cases
where the WD is simply very cool, but also (b) magnetic WDs of
suitably warmer effective temperature but with smaller radii.  These
need to be reobserved in the blue with a spectrograph and telescope of
large aperture. We made no attempt to measure the DB WDs because we
lack viable magnetic DB WD models; however, all of the DB spectra
matched well to non--magnetic DB WD models, so we believe it is
unlikely that any of the WDs in these pairs are magnetic. We could not
measure the pairs with DC WDs because there are no features with which
we can detect a magnetic field and therefore cannot rule out magnetism
without employing polarimetry or other methods of identifying a
magnetic field in these objects.

Of the remaining binary systems, we find only two that may
contain WDs with weak magnetic fields. Our automatic detection methods
are sensitive to magnetic fields between 3 MG $\leq B_{\rm WD} \leq$
30 MG; field strengths larger than this are easily identified by
visual inspection.  Therefore, there is a significant shortage of
close binary systems that could be the progenitors of the large
Intermediate--Polar and Polar CV populations.

As mentioned in \S1, \citet{Schmidt05b} discuss six newly identified
low accretion rate magnetic binary systems as being the probable
progenitors to magnetic CVs.  The magnetic field strengths of the WDs
in these systems are fairly high, with most around 60 MG.  These
objects are clearly pre-Polars and provide an obvious link between
post--common envelope, detached binaries and Polars. The existence of
these objects, however only adds to the mystery.  If observations of
these objects are possible then why have no detached binary systems
with large magnetic field WDs been detected?

Perhaps selection effects are to blame.  \citet{Schmidt05b} discuss
the various selection effects associated with targeting these
pre--Polars with the SDSS.  As is the case with the majority of the
close binary systems, the pre--Polars were targeted by the SDSS QSO
targeting pipeline \citep{Richards} which accounts for the narrow
range of magnetic field strengths found in these objects.  In the case
of significantly lower or higher magnetic field strengths, the
pre--Polars resemble an ordinary WD~+~M dwarf binary in color--color
space and are rejected by the QSO targeting algorithm.  It is possible
that this selection effect accounts for the lack of close binary
magnetic systems targeted by the SDSS as well.  Arguing against this
explanation is the large number of detached close binary systems in
our sample, and the fact that the pre--Polars were observed by the
SDSS. It is quite surprising that a detached system with a WD magnetic
field in the range required to detect these pre--Polars has not been
observed, if such objects exist.

Another selection effect discussed by \citet{Liebert05} argues that
magnetic WDs, on average, are more massive than non--magnetic WDs;
this implies smaller WD radii and therefore less luminous WDs.  Faint,
massive WDs in competition with the flux from an M star companion
might go undetected in an optical survey because they are hidden by
the more luminous, non--degenerate companion.  This would imply an
unusually small mass ratio (q = M2/M1) for the initial binary if the
progenitor of the magnetic white dwarf were massive (3-8 M$\odot$).
Thus, the magnetics may usually have been paired with an A-G star.
However, the vast majority of polars and intermediate polars with
strongly magnetic primaries have M dwarf companions.  Perhaps they
were whittled down from more massive stars by mass transfer.  The
LARPS are selected for spectroscopy because of their peculiar colors,
which arise because of the isolated cyclotron harmonics.  As
\citet{Schmidt05b,Schmidt07} point out, the WDs in LARPS are generally
rather faint (cool) and, in one case, undetected. So the large
mass/small radius selection effect would also apply to the pre--Polars
which have been observed by SDSS.

\section{Conclusions \label{sec-conclusions}}

We present a new sample of close binary systems through the Data
Release Five of the SDSS. This catalog includes more than 1200 WD~+~M
dwarf binary systems and represents the largest catalog of its kind to
date.

We have fit magnetic DA WD models \citep[see][and references
therein]{Schmidt03} to the 1100 DA WD~+~M dwarf close binaries in the
DR5 sample.  Only two have been found to potentially harbor a magnetic
DA WD of low ($B_{\rm WD} < 10$ MG) magnetic field strength.  Neither
of these potential magnetic WDs are convincing cases, though
follow--up spectroscopy to improve the S/N or polarimetry on these
objects should be performed to completely rule out the presence of a
magnetic field.

The remaining $\sim 100$ close binaries comprised of M dwarfs with
excess blue flux (:+dM) and binaries with non--DA WDs require other
means of detecting magnetic fields.  Methods that are sensitive to
magnetic fields weaker than 3 MG should also be employed on this
sample to detect possible Intermediate--Polar progenitors that may
have escaped detection with our methods.

Even if future spectroscopic or polarimetric observations reveal the
two DA WD candidates to be magnetic, their field strengths will likely
prove to be quite low.  A sample of two, detached, low magnetic field
WD binaries is not representative of the majority of known magnetic
WDs in CVs nor would it comprise an adequate progenitor population for
the newly discovered magnetic pre--Polars described in
\citet{Schmidt05b}. The question of where the progenitors to magnetic
CVs are remains unanswered by the current spectroscopically identified
close binary population.


\acknowledgments

This work was supported by NSF Grant AST 02--05875 (NMS, SLH), a
University of Washington undergraduate research grant (MPL), NSF grant
AST 03--06080 (GDS), and NSF grant AST 03--07321 (JL).

Funding for the SDSS and SDSS--II has been provided by the Alfred
P. Sloan Foundation, the Participating Institutions, the National
Science Foundation, the U.S. Department of Energy, the National
Aeronautics and Space Administration, the Japanese Monbukagakusho, the
Max Planck Society, and the Higher Education Funding Council for
England. The SDSS Web Site is http://www.sdss.org/.

The SDSS is managed by the Astrophysical Research Consortium for the
Participating Institutions. The Participating Institutions are the
American Museum of Natural History, Astrophysical Institute Potsdam,
University of Basel, University of Cambridge, Case Western Reserve
University, University of Chicago, Drexel University, Fermilab, the
Institute for Advanced Study, the Japan Participation Group, Johns
Hopkins University, the Joint Institute for Nuclear Astrophysics, the
Kavli Institute for Particle Astrophysics and Cosmology, the Korean
Scientist Group, the Chinese Academy of Sciences (LAMOST), Los Alamos
National Laboratory, the Max--Planck--Institute for Astronomy (MPIA),
the Max--Planck--Institute for Astrophysics (MPA), New Mexico State
University, Ohio State University, University of Pittsburgh,
University of Portsmouth, Princeton University, the United States
Naval Observatory, and the University of Washington.

\clearpage


\clearpage

\begin{deluxetable}{cccclcccccccccccccccccll}
\tabletypesize{\tiny}
\setlength{\tabcolsep}{0.04in} 
\tablecolumns{24} 
\tablewidth{0pt} 
\tablecaption{The SDSS--I DR5
Catalog of Close Binary Systems. \label{wdmcat}} 
\tablehead{
\colhead{Identifier} & \colhead{Plate} & \colhead{FiberID} &
\colhead{MJD} & \colhead{Sp1+Sp2\tablenotemark{a}} &
\colhead{R.A.\tablenotemark{b}} & \colhead{Decl.} & \colhead{$u_{\rm
psf}$} & \colhead{$\sigma_{u}$} & \colhead{A$_{u}$} & \colhead{$g_{\rm
psf}$} & \colhead{$\sigma_{g}$} & \colhead{A$_{g}$} & \colhead{$r_{\rm
psf}$} & \colhead{$\sigma_{r}$} & \colhead{A$_{r}$} & \colhead{$i_{\rm
psf}$} & \colhead{$\sigma_{i}$} & \colhead{A$_{i}$} & \colhead{$z_{\rm
psf}$} & \colhead{$\sigma_{z}$} & \colhead{A$_{z}$} &
\colhead{Refs\tablenotemark{c}} & \colhead{Notes\tablenotemark{d}}
\\ 
\colhead{(SDSS J)} & \colhead{} & \colhead{} & \colhead{} &
\colhead{} & \colhead{(deg)} & \colhead{(deg)} & \colhead{} &
\colhead{} & \colhead{} & \colhead{} & \colhead{} & \colhead{} &
\colhead{} & \colhead{} & \colhead{} & \colhead{} & \colhead{} &
\colhead{} & \colhead{} & \colhead{} & \colhead{} & \colhead{} &
\colhead{} \\ \colhead{(1)} & \colhead{(2)} & \colhead{(3)} &
\colhead{(4)} & \colhead{(5)} & \colhead{(6)} & \colhead{(7)} &
\colhead{(8)} & \colhead{(9)} & \colhead{(10)} & \colhead{(11)} &
\colhead{(12)} & \colhead{(13)} & \colhead{(14)} & \colhead{(15)} &
\colhead{(16)} & \colhead{(17)} & \colhead{(18)} & \colhead{(19)} &
\colhead{(20)} & \colhead{(21)} & \colhead{(22)} & \colhead{(23)} &
\colhead{(24)}} 
\startdata 
001029.87+003126.2 & 0388 & 545 & 51793 &
DZ:+dM & \phn \phn 2.62448 & \phn 00.52396 & 21.93 & 0.19 & 0.14 &
20.85 & 0.04 & 0.10 & 19.98 & 0.03 & 0.08 & 19.00 & 0.02 & 0.06 &
18.42 & 0.04 & 0.04 & EDR & \\ 001726.63$-$002451.2 & 0687 & 153 &
52518 & DA+dMe & \phn \phn 4.36099 & $-$00.41422 & 19.68 & 0.04 & 0.14
& 19.29 & 0.03 & 0.10 & 19.03 & 0.02 & 0.07 & 18.19 & 0.02 & 0.06 &
17.54 & 0.03 & 0.04 & R03 & \\ 001733.59+004030.4 & 0389 & 614 & 51795
& DA+dM & \phn \phn 4.38996 & \phn 00.67511 & 22.10 & 0.40 & 0.13 &
20.79 & 0.14 & 0.10 & 19.59 & 0.03 & 0.07 & 18.17 & 0.02 & 0.05 &
17.39 & 0.02 & 0.04 & EDR/R03 & \\ 001749.24$-$000955.3 & 0389 & 112 &
51795 & DA+dMe & \phn \phn 4.45519 & $-$00.16539 & 16.57 & 0.02 & 0.13
& 16.87 & 0.02 & 0.10 & 17.03 & 0.01 & 0.07 & 16.78 & 0.01 & 0.05 &
16.47 & 0.02 & 0.04 & EDR/R03 & \\ 002620.41+144409.5 & 0753 & 079 &
52233 & DA+dMe & \phn \phn 6.58505 & \phn 14.73597 & 17.57 & 0.01 &
0.27 & 17.35 & 0.01 & 0.20 & 17.34 & 0.02 & 0.15 & 16.65 & 0.01 & 0.11
& 16.04 & 0.02 & 0.08 & DR2 & \\ 
\enddata
\tablecomments{Table~\ref{wdmcat} is published in its entirety in the
electronic edition of the \aj. A portion is shown here for guidance
regarding its form and content. $ugriz$ photometry has not been
corrected for Galactic extinction.}  

\tablenotetext{a}{Sp1: Spectral type of the WD, Sp2: Spectral type of
the low--mass dwarf \citep[see][for details on Sp determination]{S06};
e: emission detected visually.}

\tablenotetext{b}{R.A. and Decl. are J2000.0 equinox.}

\tablenotetext{c}{EDR: \citet{Stoughton02}; DR[1,2,3]:
\citet{ADR1,ADR2,ADR3}; DR[4,5]: \citet{AMDR4,AMDR5}; R03: published
in \citet{Raymond03}; K04: published in \citet{Kleinman04}; B05:
published in \citet{van05}; Sc05: published in \citet{Schmidt05a};
S06: published in \citet{S06}; E06: published in \citet{Eisenstein06};
P05: published in \citet{Pourbaix05}; KM: published in \citet{KM06};
SN: published in \citet{Schuh06}; da06: R. da Silva (priv. comm.,
2006).}

\tablenotetext{d}{low: potential low gravity (log $g < 7$) white
dwarf.}
\end{deluxetable}

\clearpage

\begin{deluxetable}{ccccccccccccccc}
\tabletypesize{\tiny}
\setlength{\tabcolsep}{0.04in}
\tablecolumns{15}
\tablewidth{0pt}
\tablecaption{Two Potential Magnetic White Dwarf Binary Systems. 
\label{maybemags}}
\tablehead{ 
\colhead{Identifier} & \colhead{Plate} & \colhead{Fiber} & 
\colhead{MJD} & \colhead{R.A.}    & \colhead{Decl.} & \colhead{$B$} & 
\colhead{$i$} & \colhead{Sp1+Sp2} & \colhead{$u$}   & \colhead{$g$} & 
\colhead{$r$} & \colhead{$i$}     & \colhead{$z$}   & \colhead{Release} \\
\colhead{SDSS J} & \colhead{} & \colhead{} & \colhead{} & 
\colhead{(deg)} & \colhead{(deg)} & \colhead{(MG)} & \colhead{(deg)} &
\colhead{} & \colhead{} & \colhead{} & \colhead{} & \colhead{} & 
\colhead{} & \colhead{} \\
\colhead{(1)}  & \colhead{(2)}  & \colhead{(3)}  & \colhead{(4)}  &
\colhead{(5)}  & \colhead{(6)}  & \colhead{(7)}  & \colhead{(8)}  &
\colhead{(9)}  & \colhead{(10)} & \colhead{(11)} & \colhead{(12)} & 
\colhead{(13)} & \colhead{(14)} & \colhead{(15)}}
\startdata
082828.18+471737.9 & 0549 & 338 & 51981 & 127.11742 & \phn +47.29387 & 
8 & 90 & DA+dM & 20.41 & 20.35 & 20.33 & 19.58 & 19.02 & DR1 \\
125250.03$-$020608.1 & 0338 & 343 & 51694 & 193.20846 & $-$02.10227 &
$3$ & 90 & DA+dM & 19.25 & 19.12 & 18.89 & 18.31 & 17.82 & DR1 \\
\enddata
\end{deluxetable}

\end{document}